\documentclass[11pt]{article}
\usepackage{color}
\usepackage[latin1]{inputenc}
\usepackage{amsfonts,amssymb, graphicx,a4,bm,bbm}
\usepackage[english]{babel}
\usepackage{amsthm,epsfig}

\newtheorem{theorem}{Theorem}
\newtheorem{proposition}[theorem]{Proposition}
\newtheorem{corollary}[theorem]{Corollary}

\newtheorem{definition}{Definition}

\definecolor{Red}{cmyk}{0,1,1,0}

\begin{document}

\def\red{\color{Red}}

\def\reff#1{(\protect\ref{#1})}

\let\a=\alpha \let\b=\beta \let\ch=\chi \let\d=\delta \let\e=\varepsilon
\let\f=\phi \let\g=\gamma \let\h=\eta    \let\k=\kappa \let\l=\lambda
\let\m=\mu \let\n=\nu \let\o=\omega    \let\p=\pi \let\ph=\varphi
\let\r=\rho \let\s=\sigma \let\t=\tau \let\th=\vartheta
\let\y=\upsilon \let\x=\xi \let\z=\zeta
\let\D=\Delta \let\F=\Phi \let\G=\Gamma \let\L=\Lambda \let\Th=\Theta
\let\O=\Omega \let\P=\Pi \let\Ps=\Psi \let\Si=\Sigma \let\X=\Xi
\let\Y=\Upsilon

\global\newcount\numsec\global\newcount\numfor
\gdef\profonditastruttura{\dp\strutbox}
\def\senondefinito#1{\expandafter\ifx\csname#1\endcsname\relax}
\def\SIA #1,#2,#3 {\senondefinito{#1#2}
\expandafter\xdef\csname #1#2\endcsname{#3} \else
\write16{???? il simbolo #2 e' gia' stato definito !!!!} \fi}
\def\etichetta(#1){(\veroparagrafo.\veraformula)
\SIA e,#1,(\veroparagrafo.\veraformula)
 \global\advance\numfor by 1
 \write16{ EQ \equ(#1) ha simbolo #1 }}
\def\etichettaa(#1){(A\veroparagrafo.\veraformula)
 \SIA e,#1,(A\veroparagrafo.\veraformula)
 \global\advance\numfor by 1\write16{ EQ \equ(#1) ha simbolo #1 }}
\def\BOZZA{\def\alato(##1){
 {\vtop to \profonditastruttura{\baselineskip
 \profonditastruttura\vss
 \rlap{\kern-\hsize\kern-1.2truecm{$\scriptstyle##1$}}}}}}
\def\alato(#1){}
\def\veroparagrafo{\number\numsec}\def\veraformula{\number\numfor}
\def\Eq(#1){\eqno{\etichetta(#1)\alato(#1)}}
\def\eq(#1){\etichetta(#1)\alato(#1)}
\def\Eqa(#1){\eqno{\etichettaa(#1)\alato(#1)}}
\def\eqa(#1){\etichettaa(#1)\alato(#1)}
\def\equ(#1){\senondefinito{e#1}$\clubsuit$#1\else\csname e#1\endcsname\fi}
\let\EQ=\Eq
\def\0{\emptyset}

\def\pp{{\bm p}}\def\pt{{\tilde{\bm p}}}


\def\\{\noindent}
\let\io=\infty

\def\VU{{\mathbb{V}}}
\def\Pr{{\mathbb{P}}}
\def\GI{{\mathbb{G}}}
\def\TT{{\mathbb{T}}}
\def\C{\mathbb{C}}
\def\A{{\mathcal A}}
\def\P{{\mathcal V}}
\def\II{{\mathcal I}}
\def\RR{{\cal R}}
\def\SS{{\cal S}}
\def\NN{{\cal N}}
\def\EE{{\cal E}}
\def\GG{{\cal G}}
\def\PP{{\cal P}}
\def\AA{{\cal A}}
\def\BB{{\cal B}}
\def\FF{{\cal F}}
\def\v{\vskip.1cm}
\def\vv{\vskip.2cm}
\def\gt{{\tilde\g}}
\def\I{{\rm I}}
\def\rfp{R^{*}}
\def\rd{R^{^{_{\rm D}}}}
\def\ffp{\varphi^{*}}
\def\ffpt{\widetilde\varphi^{*}}
\def\fd{\varphi^{^{_{\rm D}}}}
\def\fdt{\widetilde\varphi^{^{_{\rm D}}}}
\def\pfp{\Pi^{*}}
\def\pd{\Pi^{^{_{\rm D}}}}
\def\pbfp{\Pi^{*}}
\def\fbfp{{\bm\varphi}^{*}}
\def\fbd{{\bm\varphi}^{^{_{\rm D}}}}
\def\rfpt{{\widetilde R}^{*}}

\def\tende#1{\vtop{\ialign{##\crcr\rightarrowfill\crcr
              \noalign{\kern-1pt\nointerlineskip}
              \hskip3.pt${\scriptstyle #1}$\hskip3.pt\crcr}}}
\def\otto{{\kern-1.truept\leftarrow\kern-5.truept\to\kern-1.truept}}
\def\arm{{}}
\font\bigfnt=cmbx10 scaled\magstep1

\newcommand{\card}[1]{\left|#1\right|}
\newcommand{\und}[1]{\underline{#1}}
\def\1{\rlap{\mbox{\small\rm 1}}\kern.15em 1}
\def\ind#1{\1_{\{#1\}}}
\def\bydef{:=}
\def\defby{=:}
\def\buildd#1#2{\mathrel{\mathop{\kern 0pt#1}\limits_{#2}}}
\def\card#1{\left|#1\right|}
\def\proof{\noindent{\bf Proof. }}
\def\qed{ \square}
\def\trp{\mathbb{T}}
\def\trt{\mathcal{T}}
\def\Z{\mathbb{Z}}
\def\be{\begin{equation}}
\def\ee{\end{equation}}
\def\bea{\begin{eqnarray}}
\def\eea{\end{eqnarray}}

\title {Witness trees in the Moser-Tardos algorithmic Lovász Local Lemma and Penrose trees in the hard core lattice gas}

\author{Rogério Gomes Alves$^1$, Aldo Procacci$^2$  \\\\
\footnotesize{$^1$Dep. Matem\'atica-ICEB, UFOP,
Ouro Preto - MG, 35400-000 Brazil}
\\
\footnotesize{$^2$Dep. Matem\'atica-ICEx, UFMG, CP 702
Belo Horizonte - MG, 30161-970 Brazil}\\
\tiny{emails: {rgalves@iceb.ufop.br};~ {aldo@mat.ufmg.br};} \\
}
\date{}
\maketitle


\begin{abstract}
We point out a close  connection between the Moser-Tardos
algorithmic version of the Lov\'asz Local Lemma, a central tool in
probabilistic combinatorics, and the cluster expansion of the
hard core lattice gas in statistical mechanics. We  show that the
notion of witness trees given by Moser and Tardos is essentially coincident with
that of Penrose trees in the Cluster expansion scheme of the hard
core gas. Such an identification implies that the Moser Tardos
algorithm  is successful  in a polynomial time if the Cluster
expansion converges.
\end{abstract}

\section{Introduction, state of art, notations and results}\numsec=1\numfor=1
\subsection{\it The Lovász Local Lemma}
The Lovász Local Lemma (LLL) is one of the most important tool in the framework of the so called probabilistic methods in combinatorics.
In its more general form (the non-symmetric version) it can be stated
as follows. Given a finite set $X$ and a collection ${\bm A}=\{A_x\}_{x\in X}$ of events  (the bad events) in some probability space, each
event $A_x$  with probability $Prob(A_x)$ to occur, let $G$ be a  graph with vertex set $X$ and edge set
such that  for each $x\in X$,
$A_x$ is independent of all the events in the $\sigma$-algebra
generated by $\{ A_y: y \in
X\backslash \Gamma^{*}_G(x) \}$, where $\Gamma_G(x)$ denotes the
vertices of $G$ adjacent to $x$ and
$\Gamma^{*}_G(x)=\Gamma_G(x)\cup\{x\}$. A graph $G$  with these characteristics is called a {\it dependency graph} for the collection of events $\bm A$.
 Let  $\bar A_x$ be  the complement
event of $A_x$ so that   $\bigcap_{x\in X}\bar A_x$ is  the event that none of the events
$\{A_x\}_{x\in X}$  occurs. The Lov\'az local lemma gives a sufficient
criterion to guarantee that $ \bigcap_{x\in X}\bar A_x$ has positive probability (and hence is non empty).

\begin{theorem}[{\bf Lov\'asz Local Lemma}]
\label{LLLold} Let  $G$ be a dependence
graph for the collection of events $\{A_x\}_{x\in X}$   with probability
${Prob}(A_x)=p_x$ and  suppose there exists a sequence $\bm{\mu}=\{\mu_x\}_{x\in X}$ of real numbers
 in $[0,+\infty)$ such that, for each $x\in X$,
$$
p_x\;\leq\;  {\mu_x \over  \prod_{y\in \Gamma^*_G(y)} (1+\m_y)}
$$
 Then
$$
{Prob}(\bigcap_{x\in X}\bar A_x)\,> \,0
$$
\end{theorem}
This lemma, originally formulated by Erd\"{o}s and Lovász in \cite{EL}, has been heavily used in combinatorics to obtain bounds on  problems
about  graph coloring, k-sat, latin transversal, Ramsey numbers, and so on. Shearer \cite{Sh}
gave an alternative formulation
of this lemma which has been used as a bridge by  Scott and Sokal \cite{SS,SS2} to point out a surprising and very interesting connection between with the cluster expansion of the hard core lattice gas on $G$ (the dependency graph). We remind rapidly below the hard core gas setting and  its state of the art.
\subsection{\it The self-repulsive Hard core gas on a graph $G=(X, E)$}
The hard core gas on  a graph  $G$ with vertex set $X$ and edge set $E$ is defined as follows.
\def\E{\mathbb{E}}
 Suppose  that each vertex $x\in X$ can be occupied by a 'particle' (also called sometimes, depending on the context, a 'polymer') or can be left empty. Moreover each  particle occupying the vertex $x\in X$
carries an ``activity" $w_x \in \mathbb{C}$ and we  denote by $\bm w =\{w_x\}_{x\in X}$ the set of all activities.
We further suppose that this gas of particles on $G$
interacts through a self repulsive hard core nearest neighbor  pair potential. Namely, each vertex can be occupied at most by one particle,
and if a particle occupies the vertex $x\in X$, then all neighbor vertices of $x$ in  $G$ must be empty.
In the statistical mechanics lingo, if $x,y$ are  vertices of the graph $G=(X,E)$ where the hard core gas is defined  such that either $\{x,y\}\in E$ or $x=y$, it is usual to say that
{\it $x$ and $y$ are incompatible} and write $x\not\sim y$ (and compatible otherwise, i.e. if $\{x,y\}\not\in E$ and $x\neq y$, writing $x\sim y$).
The grand-canonical partition function of this gas  in the ``volume" $X$  is then defined as
$$
\Xi_{X}(\bm w)=\sum_{Y\subset X \atop Y ~{\rm independent}}
\prod_{y\in Y}w_y \Eq(partiz)
$$
where the sum in the r.h.s. is over the independent subsets of the vertex set $X$ of $G$ (a subset $Y\subset X$ is independent in $G$ if no edge of $G$ has both
endpoints in $Y$) so that $\Xi_{X}(\bm z)$ coincides with the independent set (multivariable) polynomial on $G$.
The  ``pressure'' of this gas is defined via the formula (hereafter, whenever $X$ denotes a set, $|X|$ denotes its cardinality)
$$
P({\bm w})={1\over |X|}\log \Xi_{X}({\bm w})\Eq(pressure)
$$
Moreover  another key function   is
$$
\Pi_{x_0}(\bm w)={\partial\over \partial w_{x_0}}\log \Xi_{X}(\bm w)= { \Xi_{X\setminus \G^*_G(x)}(\bm w)\over  \Xi_{X}(\bm w)}
\Eq(relpilog)
$$
The quantity $w_x \Pi_{x_0}(\bm w)$ is, from the physical point of view (at least for positive activities  $\bm w\ge 0$) the one-point correlation function of the hard core gas
(i.e. the probability to see a particle sitting in the site $x_0$ regardless of where the other particles are).

It is a well known fact that $\log \Xi_{X}({\bm w})$ (and hence  $P({\bm w})$ and $\Pi_{x_0}(\bm w)$)  can be written in term of a formal series, known as
{\it cluster expansion} (CE) of the hard-core gas. Indeed, let  $G_n$  denote the set of all connected graphs with vertex set $\I_n=\{1,2,\dots, n\}$ and,
given an $n$-tuple  $(x_{1},\dots ,x_{n})\in X^n$, let $g(x_1 ,\dots ,x_n)$  be the graph with vertex set $\I_n$
which has the edge $\{i,j\}$
if and only if $x_i\not\sim x_j$.
Define, for $n\ge 2$
$$
\phi^{T}(x_{1},\dots ,x_{n})=\cases{
\sum\limits_{g\in {G}_n\atop
g\subset g(x_1 ,\dots ,x_n)}(-1)^{|E_g|}& if
$g(x_1 ,\dots ,x_n)\in G_n$\cr\cr
~~~~~~0 & if $g(x_1 ,\dots ,x_n)\not\in G_n$
}
\Eq(7)
$$
Then, one can write formally  (see e.g. \cite{C,FP,pdls,PS, So})
$$
\log\Xi_{X}({\bm w})= \sum_{n=1}^{\infty}{1\over n!}
\sum_{(x_{1},\dots ,x_{n})\in X^n}
\phi^{T}(x_1 ,\dots , x_n)\,w_{x_1}\dots\,w_{x_n}\Eq(6)
$$
whence
$$
\Pi_{x_0}(\bm w)=\sum_{n=0}^{\infty}{1\over n!} \sum_{(x_1,\dots,x_n)\in X^n}
\phi^{T}(x_0,x_1 ,\dots , x_n){w_{x_1}}\dots{w_{x_n}}\Eq(PiLa)
$$

The equations \equ(6) and \equ(PiLa)  make sense  only for those $\bm w\in \C^{|X|}$ such that  the formal series in the r.h.s. of \equ(6) and \equ(PiLa) converge absolutely.
It is again a well known fact (see, e.g. \cite{Ru} and also Proposition \ref{pro9} ahead) that the number $\phi^{T}(x_{1},\dots ,x_{n})$ defined in \equ(7) has the following property
$$
\phi^{T}(x_{1},\dots ,x_{n})=(-1)^{n-1}|\phi^{T}(x_{1},\dots ,x_{n})|\Eq(alternate)
$$
We can thus consider, for $\bm \r=\{\r_x\}_{x\in X}$ with  $\r_x\in (0,\infty)$ for all $x\in X$, the positive term series
$$
\Pi_{x_0}(-\bm \r)= \sum_{n=0}^{\infty}{1\over n!} \sum_{(x_1,x_2,\dots,x_n)\in X^n}
|\phi^{T}(x_0,x_1 ,\dots , x_n)|\,\r_{x_1}\cdots{\r_{x_n}}\Eq(TP)
$$
and, if we are able to show that $\Pi_{\g_0}(-\bm \r)$ converges
for some (bounded) positive value  $\bm\r\in [0,\infty)^{|X|}$, then also $\Pi_{x_0}(\bm w)$
converges absolutely, whenever $\bm w=\{w_x\}_{x\in X}$ is in the poly-disk $\{|w_x|\le \r_x\}_{x\in X}$ and in this poly-disk the pressure
\equ(pressure) admits
the bound uniform in $X$
$$
|P(\bm w)|\le \sup_{x_0\in X}\,\,\r_{x_0}\,\Pi_{x_0}(-\bm \r)
$$
Throughout this paper, operations and relations involving boldface
symbols should be understood componentwisely, for instance $|\bm w|\le \bm \r$ indicates shortly $\{|w_x|\le\r_x\}_{x\in X}$ and  $-\bm w$ means $ \{-w_x\}_{x\in X}$, etc..
 Note that the region $|\bm w|\le \bm \r$  is also a zero-free region of the partition function $\Xi_{X}(\bm w)$.

 The set
 $$
 {\cal R}(G)=\{\bm \r\in [0,\infty)^{|X|}: ~ \Pi_{x_0}(-\bm \r)<+\infty\}
 $$
constitutes, in the statistical mechanics lingo,
the {\it convergence region} of the cluster expansion. Observe that, by definition, ${\cal R}(G)$ is a down-set, i.e.
 $\bm \r\in {\cal R}(G)$ and $\bm \r'\le \bm \r$ implies $\bm \r'\in {\cal R}(G)$.

A lot of efforts has been employed during the past three decades
 to establish efficient upper bounds  for  ${\cal R}(G)$ (see e.g. \cite{GK, C, Se, KP, PS}). These efforts can be resumed by the so called
Dobrushin criterion \cite{D}, which can be stated as follows.

\begin{theorem}[{\bf Dobrushin}]\label{dob}  Let   $G=(X, E)$ be a graph let ${\cal R}(G)$  the {\it convergence region} of the cluster
expansion of the hard core gas on $G$. Let $\bm{\mu}=(\mu_x)_{x\in X}$ be a family of non negative numbers in
$[0,+\infty)$ and let $\bm{\r}=(\r_x)_{x\in X}$  such that, for all $x\in X$
$$
\r_x\le {\mu_x \over  \prod_{y\in \Gamma^*_G(y)} (1+\m_y)}
$$
Then $\bm\r\in {\cal R}(G)$ and
$$
\r_{x}\Pi_x(-\bm \r)\le \m_x
$$
\end{theorem}
\\In 2007 however the Dobrushin criterion
has been improved by Fernández and Procacci \cite{FP}.

\begin{theorem}[{\bf Fernández-Procacci}]\label{FP}  Let   $G=(X, E)$ be a graph let ${\cal R}(G)$  the {\it convergence region} of the cluster expansion of the hard core gas on $G$. Let $\bm{\mu}=(\mu_x)_{x\in X}$ be a family of non negative numbers in
$[0,+\infty)$ and let $\bm{\r}=(\r_x)_{x\in X}$ such that, for all $x\in X$
$$
\r_x\le {\mu_x \over \sum\limits_{R\subseteq  \Gamma^*_G(x)\atop R\ {\rm indep\ in}\ G}
\prod_{x\in R}\mu_x}
$$
Then $\bm \r\in {\cal R}(G)$ and
$$
\r_{x}\Pi_x(-\bm \r)\le \m_x
$$
\end{theorem}
\\The improvement respect to Theorem \ref{dob} is immediately recognized by noting that
$$
\prod_{y\in \Gamma^*_G(y)} (1+\m_y)= \sum\limits_{R\subseteq  \Gamma^*_G(x)}
\prod_{x\in R}\mu_x \ge \sum\limits_{R\subseteq  \Gamma^*_G(x)\atop R\ {\rm indep\ in}\ G}
\prod_{x\in R}\mu_x
$$
\subsection{\it Connection between LLL and hard core gas}

As anticipated  above, in 2005  Scott and Sokal \cite{SS,SS2} elucidated a surprising and very interesting connection between
the repulsive hard core gas in statistical mechanics and the Lovász local lemma in probabilistic combinatorics. In particular,
they  pointed out  that the Shearer formulation \cite{Sh} for the applicability of the LLL was equivalent
to require the convergence of the cluster expansion of the hard core lattice gas. As an immediate consequence, they showed
that the LLL condition of Theorem \ref{LLLold} could be seen as  a reformulation of  the  Dobrushin criterion of Theorem \ref{dob} for the
convergence of the cluster expansion.
Scott and Sokal reformulated the Shearer version of the LLL in terms of convergence of the cluster expansion of the hard core gas as follows.

\begin{theorem} [{\bf Scott-Sokal}]\label{elo}
Let
 $G$ be  a dependence
graph for the family of events  $\{A_x\}_{x\in X}$ with probability
${Prob}(A_x)=p_x$. Let $ \Xi_X(\bm w)$ be the  partition function of the hard core  gas  on $G$ and let ${\cal R}(G)$  the {\it convergence region} of the cluster expansion of the hard core gas on $G$.

\\If $\bm p=\{p_x\}_{x\in X}\in {\cal R}(G)$,  then,
$$
Prob(\bigcap_{x\in X}\bar A_x)\;\ge\; \Xi_X(-\bm p)\;>\; 0\;.\Eq(pbc)
$$
Furthermore these bounds
are  the best possible, i.e. if $\bm p\notin {\cal R}(G)$, then there can be constructed a family of
events $\{B_x\}_{x\in X}$  in a suitable probability space with probabilities ${Prob}(B_x) = p_x$ and  dependency
graph $G$, such that $\mathbb{P}(\bigcap_{x\in X}\bar B_x)=  0$.

\end{theorem}
\\{\bf Remark}. By merging Theorem \ref{dob} into Theorem \ref{elo} one obtains immediately the usual LLL, i.e. Theorem \ref{LLLold}. On the other hand, by
merging Theorem \ref{FP} into Theorem \ref{elo} we have immediately
the improved version of the LLL recently given by Bissacot et al. \cite{BFPS}.

\begin{theorem}[{\bf Bissacot-Fernández-Procacci-Scoppola}]\label{LLL} Let  $G$ be a dependence
graph for the collection of events $\{A_x\}_{x\in X}$   with probability
${Prob}(A_x)=p_x$ and  suppose there exists a sequence $\bm{\mu}=(\mu_x)_{x\in X}$ of real numbers
 in $[0,+\infty)$ such that, for each $x\in X$
$$
              p_x\;\leq {\mu_x \over \sum\limits_{R\subseteq  \Gamma^*_G(x)\atop R\ {\rm indep\ in}\ G}
\prod_{x\in R}\mu_x}\Eq(FP)
$$
Then
\def\A{{\bm A}}
\begin{eqnarray}\nonumber
\mathbb{P}(\bigcap_{x\in X}\bar A_x) ~>~0
\end{eqnarray}
\end{theorem}
\\Theorem \ref{LLL} has been already used in \cite{NPS} and \cite{BKP} to obtain improved bounds  on various graph coloring problems.
\subsection{\it The Algorithmic Moser-Tardos version of the Lovász Local Lemma}
The  unquestionable popularity of the LLL came however always together with a criticism about its inherent non-constructive character. Namely the LLL,  giving sufficient conditions for the probability that
none of the undesirable events occur to be strictly positive, implies that there exist at last one  configuration in the probability space of the events
which realizes the occurrence of the "good" event
$\bigcap_{x\in X}\bar A_x$ , but it   does not provide any algorithm capable to produce, possibly in a polynomial time,
such a configuration. Efforts to find an  algorithmic version of the LLL go back to the work of Beck
 \cite{B} and Alon \cite{A}, and,  after various contributions  (see e.g. \cite{Mo} and references therein),  finally culminate in a recent  breakthrough  paper by Moser and Tardos \cite{MT}, who  gave a fully  algorithmic version of LLL if the events are restricted to a class which however covers basically  all known applications of LLL. The Moser Tardos scheme is as follows.
Let $\P$ be a finite family of mutually independent random variables. Let $X$ be a finite set and let  $\bm A=\{A_x\}_{x\in X}$ be
a finite family of events, each $A_x$  depending  by some subset of the random variables  of the family $\P$, each with probability
${Prob}(A_x)=p_x$. Denote $vbl(A_x)$, for all $A_x \in \bm A$, the minimal (with respect to inclusion) and unique subset of $\P$ that determine $A_x$.
The dependence graph of the family $\bm A$ is the graph $G=(X,E)$  with vertex set $X$ and edge set $E$ is constituted by the pairs
$\{x,x'\}\subset X$  such that $vbl(A_x) \cap vbl(A_{x'}) \neq \emptyset$.   Observe that if $x,y\in X$ and $vbl(A_x) \cap vbl(A_{y}) \neq \emptyset$ then
either $\{x,y\}\in E$ or $x=y$. By analogy with the hard core gas we  denote this    with the symbol  $x\not\sim x'$ and say that
$x,  y$ {\it are incompatible} or {\it overlap} (so $x$, $y$ compatible, denoted with $x\sim y$ means $\{x,y\}\not\in E$, i.e. $vbl(A_x) \cap vbl(A_{x'}) = \emptyset$). Within this scheme
 Moser and Tardos defined the following algorithm.
\vv
\\{\it MT-Algorithm}.  As initial step choose a  random evaluation of the variables $\n\in \P$. If some $ A\in \bm A$ occurs, then  pick one  of them
   (at random or according to some
   deterministic rule), say $A_x$
   and  take a new evaluation (resampling) only of its variables, keeping unchanged all the other variables in $\P$.
   The algorithm stops when we reach an evaluation of the variables $P\in \P$ such that none of the
   events in the family $\bm A$ occurs.
The first step of the algorithm is the initial sampling of all variables in $\P$ (the step $0$ by convention)
   and for $i\in \mathbb{N}$, the step $i$ of the algorithm is the selection (according to some deterministic or random rule)
   of an occurring bad event $A_x\in \bm A$ and  the resampling of its  variables $vbl(A_x)$.

\begin{theorem}[{\bf Moser Tardos}]\label{MosTar}
Let $\P$ be a finite set of mutually independent random variables. Let $\bm A=\{A_x\}_{X}$
be a finite set of events determined by these variables, each with probability $Prob(A_x)=p_x$ and  with dependency graph $G$. Suppose there exists a sequence $\bm{\mu}=(\mu_x)_{x\in X}$ of real numbers
in $[0,+\infty)$ such that, for each $x\in X$
$$
p_x\;\leq\;  {\mu_x \over  \prod_{y\in \Gamma^*_G(y)} (1+\m_y)}\Eq(mta)
$$
Then there exists an assignment of values to the variables $\P$ such that none  of the events in $\bm A$ occurs. Moreover
the randomized algorithm described above resamples an event $A_x\in \bm A$, at most an expected $\m_x$
times before it finds such an evaluation. Thus the expected total number of resampling  steps is at most
$\sum_{x\in X}\m_x$.
\end{theorem}

Following Moser and Tardos, as  the algorithm runs, resampling at each step some bad event from the family $\bm A$, one can define
{\it the Log of the algorithm $ C=\{C(1),C(2),\dots\}$} with $C(i)\in X$. Namely, $ C$ lists the events as they are selected and resampled by the algorithm at each step,
so that, for  $i\in \mathbb{N}$, if $C(i)=x$ then the event $A_x\in \bm A$ is picked and resampled at step $i$ of the algorithm.
  Note that if the algorithm stops then $C$ is partial, i.e. there exists an $n\in \mathbb{N}$
  such that $C: \I_n\to X$. Using the words of Moser Tardos, $C$  is a random variable determined by the random choices made by the algorithm at each step.

\vv
\\{\it Rooted trees, dressed trees and witness trees}. Moser and Tardos' proof of Theorem \ref{MosTar} is based on the notion of a `witness tree'. To explain these objets we need to
introduce some notations and definitions about trees.

An undirected unlabeled simple connected graph with  no cycles  and such that one vertex has been designated the root is called a {\it rooted tree}. Hereafter we will use the letter
$t$ to denote a  generic unlabeled rooted tree and we denote by $\Theta$ the set of all possible unlabeled rooted trees.
A rooted tree $t\in \Theta$ has a natural partial order (called the tree order). Namely, given two (distinct) vertices $u$ e $v$  in a  rooted tree,
 $v$ is said to   be a {\it descendant} of $u$ or ($u$ s an {\it ancestor} of $v$), if there is a path from the root  to $v$ which  contains $u$.
If $\{v, u\}$
is an edge of a rooted tree, then either $v$   is a descendant of $u$  or viceversa. So actually  any edge $\{u,v\}$
in a rooted tree is directed (i.e. is an ordered pair) and we write $(u,v)$ ($v$ descendent of $u$) with
$u$ being the {\it parent} (or predecessor, or father) and $v$ being  the {\it child} (or successor). Note that
each vertex in $t$ different form the root has one and only one parent.
The root has no predecessor and it is the extremum respect to the partial order relation.

Given a set $X$,
a {\it $X$-dressed  tree} is a pair $\t=(t,\s)$ where $t\in \Th$ is a rooted tree  with vertex set $V_t$ and $\s$ is function $\s: V_t\to X $. Note that with this definition a dressed
tree with labels in $X$ may have distinct vertices associated to the same label in $X$, i.e. the function $\s$ may not be an injection.

According to Moser and Tardos, the definition of witness trees is as follows.
\vv

\begin{definition}\label{pippo} Let $X$ be the vertex set of a graph $G$. A  witness tree $\t = (t, \s)$ is a $X$-dressed tree with  $t\in \Th$ and
$\s: V_t\to X$ such that the children of a vertex  $u\in V_t$ receive labels from $\G^*(\s(u))$ and these labels are  distinct.
\end{definition}
\vv
\\Moser Tardos then associate to each step $s$ of the algorithm, with log $C$,  a witness  tree $\t(s)$
(with root labeled $C(s)$) via a well defined iterative process. We will see in  section 3 the details of this process. Let us
denote by ${\cal T}^{x_0}_{X}$ the set of all distinct witness trees $\t=(t,\s)$ that can be obtained
via the algorithm and are such that $t$ is a rooted tree with root labeled $x_0$.

\\Moser and Tardos then prove (lemma 2.1 in \cite{MT}) that the probability $Prob(\t)$ that  a witness tree $\t=(t,\s)$ with vertex set $V_t$ and labels $\{\s(v)\}_{v\in V_t}$ at the vertices appears in the log $C$ of the algorithm is at most
$$Prob(\t)\le ~\prod_{v\in V_t} { Prob}(A_{\s(v)})~ \equiv ~\prod_{v\in V_\t} p_{\s(v)}\Eq(bound)$$
Now,  for $x\in X$ let  $N_x$ be the random variable that counts how many times the event
$A_x$ is resampled during the execution of the MT-algorithm.
Then $N_x$ is, by definition, the number of occurrences of the event $A_x$ in the log $C$ of the algorithm and also the number of {\it distinct} proper witness trees
occurring in $C$ that have their root labeled $x$.  {\it Therefore one can bound the expectation of $N_x$ simply by summing the bounds \equ(bound)
on the probabilities $Prob(\t)$ as $\t$ varies in the set ${\cal T}^{x}_X$  of the different  witness trees with root labeled $x$}. Thus the expected value $E(N_x)$ of $N_x$ is bounded as
$$
E(N_x)\le \Phi_x(\bm p)\Eq(expn)
$$
where
$$
\Phi_x(\bm p)=
\sum_{(t,\s) \in {\cal T}^{x}_X} \prod_{v\in V_t}p_{\s(v)}\Eq(77)
$$
Moser and Tardos's conclude their proof by  showing, via a Galton-Watson branching process argument, that the quantity $\Phi_x(\bm p)$ defined in \equ(77)  is bounded by $\m_x$ if
 probabilities $\{p_x\}_{x\in X}$ are such that conditions \equ(mta) are verified.

\subsection{\it Results}
Afterwards the work of Scott and Sokal, relating the non-constructive Lovász Local Lemma to the statistical mechanics of hard core gas and the consequent improvement
of the lemma by Bissacort et al. obtained exploiting this connection, it is a natural question to ask whether there can be made a similar connection
between the algorithmic Lovász Local Lemma (ALLL) proposed by Moser and Tardos and the hard core gas. We stress that  question is far from being trivial, since the scheme proposed by Moser
and Tardos to prove their Theorem \ref{MosTar}, based on the concept of witness trees, has, at first sight, nothing to do with
the  various proofs  of the non-algorithmic Lovász Local Lemma proposed in the literature.

Strong indications that a connection between the  ALLL and the  hard core gas must indeed exist come form
two recent works \cite{KS} and \cite{Pe}. In \cite{KS}   Kolipaka and Szegedy relate  the Moser Tardos algorithm
to  the set of Shearer conditions via an auxilary algorithm (called by the authors ``generalized resample") and  a reformulation of the Moser Tardos scheme in which
the notion of witness trees is replaced by two  alternative concepts (called by the authors ``stable set sequences" and ``stable set matrices").
However, in \cite{KS}
no explicit improvement on Theorem \ref{MosTar} eventually based on the equivalence of Shearer conditions and convergence of the cluster expansion is presented.
The improvement was later found by Pegden \cite{Pe} using a completely different method. Namely, Pegden realized that, within the Moser Tardos scheme
involving witness trees, it was possible to modify  the branching process argument given in \cite{MT} in order
to adapt it to the Bissacot et al. condition  \cite{BFPS} of Theorem \ref{LLL}.

In this paper we show that the   connection between ALLL and CE of the hard core gas is astonishingly direct,  much more direct, dare we say,
than the one pointed out by Scott and Sokal  for the non-constructive LLL. Indeed, the connection can be obtained
bypassing completely Shearer formulation and  remaining within  the original Moser Tardos scheme involving witness trees (as the work of Pegden was implicitly suggesting).
Namely, by a slight modification of the  map which defines the Penrose trees in CE, we are able to  show that  the  notion of witness tree defined in \cite{MT}
is  in fact coincident with that of the Penrose tree in the CE scheme of the hard core gas.
Such an identification  implies that the sum over witness trees given in \cite{MT}, which bounds from above the expected number of steps an event $A_x$ is resampled, happens to be
exactly equal to the cluster expansion  of the one point correlation function defined in  \equ(relpilog) calculated at $\bm w= -\bm p$ (we recall that $\bm p=\{p_x\}$ with
$p_x\in [0,1]$ being the  probability
$P(A_x)$ of occurrence of the event $A_x$).
The main result of the paper can be resumed by the following theorem.

\begin{theorem}\label{MTn}
Let $\P$ be a finite set of mutually independent random variables in
a probability space. Let $\bm A=\{A_x\}_{x\in X}$ be a finite set of
events determined by these variables, each with probability
$Prob(A_x)=p_x$, with dependency graph $G$. Let $\Xi_X(\bm w)$ be the
partition function of the hard-core lattice gas on $G$ with
complex activities ${\bm w}=\{w_x\}_{x\in {X}}$ and let ${\cal R}(G)$  the {\it convergence region} of the cluster expansion of the hard core gas on $G$.

{ If $\bm p=\{p_x\}_{x\in X}\in {\cal R}(G)$}, then there exists an assignment of values to the variables $\P$ such that none  of the events in $\cal A$ occurs. Moreover
the randomized algorithm described above finds such an evaluation  resampling an event $A_x\in \cal A$ in  an expected time $T_x$ such that
$$
T_x\le p_x\left[{\partial\log \Xi_X(\bm w)\over \partial w_x}\right]\Bigg|_{\bm w =-\bm p}\Eq(tx)
$$
and  the expected total number of resampling $T$ of the variables in $\cal P$  is at most
$$
T\le \sum_{x\in X}p_x\left[{\partial\log \Xi_X(\bm w)\over \partial w}\right]\Bigg|_{\bm w =-\bm p}\le |X||P(\bm w={-\bm p})|
$$
where $P(\bm w)$ is the pressure of the hard core  lattice gas on  $G$ with activities $\bm w=\{w_x\}_{x\in X}$.
\end{theorem}

\v
\\{\bf Remark 1}. Theorem \ref{MTn} above together with  Theorem \ref{FP} immediately yields for free the following corollary, which is the result obtained by Pegden.
\begin{corollary}
Under the hypothesis of Theorem \ref{MTn}, if  $\bm{\mu}=\{\mu_x\}_{x\in X}$  is a sequence  of real numbers
 in $[0,+\infty)$ such that, for each $x\in X$
$$
p_x\;\leq\;  {\mu_x \over  \sum\limits_{R\subseteq  \Gamma^*_G(x)\atop R\ {\rm indep\ in}\ G}
\prod_{y\in R}\mu_y}
$$
then
the randomized algorithm  resamples an event $A_x\in \bm A$, at most an expected $\m_x$
times before it finds such an evaluation. Thus the expected total number of resampling steps is at most
$
\sum_{x\in X}\m_x
$.
\end{corollary}
\\{\bf Remark 2}.
If $\bm p$ is outside the convergence radius of the cluster expansion, then one can say nothing about the efficiency
of the algorithm since the series
bounding the expected time the algorithm stops diverges. Of course in the algorithmic  Moser Tardos setting , i.e. the collection of  bad events $\bm A=\{A_x\}_{x\in X}$
depending on a finite number of independent random variables $\n\in {\P}$, it is possible to construct different algorithms which
could be more efficient and stops even for a set of probabilities  $\bm p$  for which the Moser Tardos algorithm doesn't stop.
Along these directions we would like to cite some interesting results obtained in \cite{GKM} and \cite{EP}.

\vv

The next two section are devoted to the proof of Theorem \ref{MTn}. Specifically,  in section 2 we define the modified Penrose
map for the CE of the hard core gas on a graph $G$ and write the series for the pressure and the derivative of the log of
the partition function in terms of a sum over
Penrose trees. In section 3 we show that the witness trees of the Moser Tardos scheme coincide with these modified Penrose trees and conclude the proof of Theorem \ref{MTn}.

\section{Cluster Expansion on the hard core gas
on a graph.  A variant of  the Penrose map }\def\E{\mathbb{E}}

\def\v{\vskip.1cm}
\def\vv{\vskip.2cm}
\def\gt{{\tilde\g}}
\def\E{{\mathcal E} }
\def\I{{\rm I}}
\def\GI{\mathbb{G}}
\def\EE{\mathbb{E}}\def\VU{\mathbb{V}}

\\We now reorganize the series $\Pi_{x_0}(-\bm\r)$ of equation \equ(TP) via Penrose map and Penrose identity.
 To this purpose, we need to recall some definitions.
In particular, a very special role in order to state the Penrose identity is played by the so-called {\it labeled rooted trees} and {\it plane rooted trees}.

\subsection{\it Labeled Trees and plane rooted tress}

We will use the following notations. Given a vertex $v\neq 0$ in a (unlabeled) rooted tree $t$, its {\it depth},
denoted by $d(v)$, is the number of edges in the unique path from the root to that vertex.
Given a vertex $v$ in a rooted tree different from the root, we denote by $v^*$ is  parent and
we denote by $s_v$ the number of its children. 
Note that $d(v^*)=d(v)-1$. Children of the same parent are also called {\it siblings}.
Given a vertex $v$ in a rooted tree, any vertex $w$ such that $d(w)=d(v)$ but $w$ is not a sibling of $v$ is called a  {\it cousin} of $v$ and
any vertex $w\neq v^*$ such that $d(w)=d(v)-1$ is called an {\it uncle} of $v$.
\vv
\\{\it Plane rooted trees}. A  plane tree is a rooted tree $t$ for which an ordering is given for the children of each vertex.
An ordering of the children in a rooted tree $t$ is equivalent to a drawing  of $t$ in the plane,
obtained,  e.g., by putting  parents at the left of their children which are ordered
in the top-to-bottom order. Note that the number of plane rooted trees with $n$ vertices is
always greater that the number of  rooted trees with $n$ vertices. E.g. there are 4 different rooted trees with 4 vertices while there are 5 different plane rooted trees
with 4 vertices. We denote by $ {\TT}^0_n$  the set of all plane rooted trees
with $n+1$ vertices.

\vv
\\{\it Labeled  rooted trees}.
Let  $\I_n^0=\{ 0,1,2,\dots,n \}$. A rooted tree $t$ with vertex set $\I^0_n$ and root $0$  is usually called a {labeled rooted tree}. In other words,
according to  the notations adopted in sec. 1.4, a labeled rooted tree is a $\I_n^0$-dressed rooted tree $\t=(t,\s)$ where $t$ is a rooted tree with vertex set $V_t$ and root $r$
and $\s:V_t\to  \I_n^0$ is a bijection  (therefore $|V_t|=n+1$) such that $\s(r)=0$.
Note that the number of labeled rooted trees with $n$ vertices is
always greater that the number of  plane rooted trees with $n$ vertices. E.g. there are 16 different labeled rooted trees with 4 vertices while there are 5 different plane rooted trees
with 4 vertices.
We will use the letter
$\th$   to denote a generic labeled tree for which the vertex  $0$ has been chosen as the root
and we denote by $T^0_{n}$ the set of all  labeled trees with vertex set
$\I_n^0$ which are rooted in $0$.

\vv
There is a natural map $m: T^0_{n}\to \mathbb{T}^0_n$
which  associates to each   labeled rooted tree $\th\in T^0_{n}$  a  unique plane rooted tree $m(\th)\in \mathbb{T}^0_n$. This unique plane rooted tree $m(\th)$
is obtained
by fixing the order of the children in each vertex of $\th$ according  with the order of their labels in $\I^0_n$.
For example the plane rooted trees associated to the trees $\th_1$ with edge set
$\{0,3\},\{1,3\}, \{2,3\}, \{2,4\}$, $\th_2$ with edge set
$\{0,2\},\{1,4\}, \{2,3\}, \{2,4\}$,   $\th_3$ with edge set $\{0,2\},\{0,3\}, \{1,3\}, \{3,4\}$ and $\th_4$ with edge set
$\{0,2\},\{0,4\}, \{2,3\}, \{1,2\}$
are drawn below.
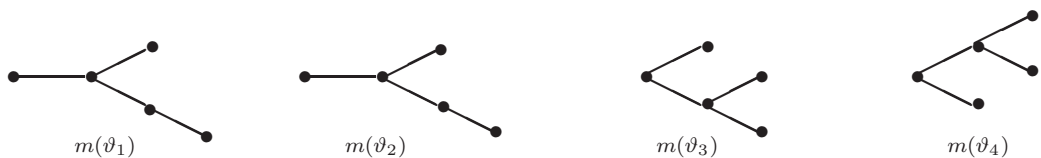
\vv
\setlength{\unitlength}{.9cm}
$$
\begin{picture}(15,2.0)
\thicklines
\put(1.0,0.5){$\scriptstyle m(\th_1)$}
\put(0.0,1.5){$\bullet$}
\put(0.15,1.61){\line(1,0){1}}
\put(1.15,1.5){$\bullet$}
\put(1.25,1.61){\line(2,1){0.8}} %
\put(2.05,1.95){$\bullet$} %
\put(1.25,1.6){\line(2,-1){0.8}}
\put(2.01,1.02){$\bullet$}
\put(2.10,1.15){\line(2,-1){0.8}}
\put(2.85,0.62){$\bullet$}

\put(5.0,0.5){$\scriptstyle m(\th_2)$}
\put(4.3,1.5){$\bullet$}
\put(4.45,1.61){\line(1,0){1}}
\put(5.45,1.5){$\bullet$}
\put(5.55,1.6){\line(2,1){0.8}} %
\put(6.31,1.91){$\bullet$} %
\put(5.55,1.6){\line(2,-1){0.8}}
\put(6.35,1.06){$\bullet$}
\put(6.40,1.2){\line(2,-1){0.8}}
\put(7.12,0.69){$\bullet$}

\put(9.6,0.5){$\scriptstyle m(\th_3)$}
\put(9.35,1.5){$\bullet$}
\put(9.45,1.65){\line(2,1){0.8}} %
\put(10.25,1.95){$\bullet$} %
\put(9.45,1.6){\line(2,-1){0.8}}
\put(10.25,1.10){$\bullet$}
\put(10.30,1.2){\line(2,-1){0.8}}
\put(10.30,1.2){\line(2,1){0.8}}
\put(11.05,0.69){$\bullet$}
\put(11.05,1.5){$\bullet$}

\put(13.9,0.5){$\scriptstyle m(\th_4)$}
\put(13.35,1.5){$\bullet$}
\put(13.45,1.65){\line(2,1){0.8}} %
\put(14.25,1.95){$\bullet$} %
\put(13.45,1.6){\line(2,-1){0.8}}
\put(14.25,1.10){$\bullet$}
\put(14.30,2.1){\line(2,-1){0.8}}
\put(14.30,2.1){\line(2,1){0.8}}
\put(15.05,2.4){$\bullet$}
\put(15.05,1.6){$\bullet$}

\end{picture}
$$
\def\T{{\mathcal{T}}}
\\Observe $\th_1$ and $\th_2$, which are different labeled trees, are sent
by the map $m$  into the same plane rooted tree, i.e. $m(\th_1)=m(\th_2)$. On the other hand $m(\th_3)$ and $m(\th_4)$ are {\it different} plane rooted trees
 (even though they correspond to the same unlabeled rooted tree).

Clearly the  map $\th\mapsto m(\th)=t$ is many-to-one and the cardinality of the
preimage $m^{-1}(t)$ of a plane rooted tree
$t$ is equal to the number of ways of labeling the $n$ non-root vertices of $t$ with $n$ distinct labels from $\{1,2,\dots, n\}$
consistently with order of the children in each vertex, i.e.,
$$
\card{\{\th\in T^0_n: m(\th)=t\}}\;=\; {n!\over\prod_{v\in V_t} s_{v_i}!}\Eq(rel1)
$$
where recall that if $v\in V_t$, then $s_v$ denotes the number of the children of $v$.


\vv

There is also a natural map $\theta: \mathbb{T}^0_n\to T^0_n$ (an injection)
which assigns to the vertices of a plane rooted tree $t$  labels in the set $\I^0_n$ in the following  natural way:
the root has label $0$, the $s_0$ children of the root have labels $1,2,\dots, s_0$ from top to bottom,
the higher root child vertex, i.e. that  with label 1, has $s_1$ children with labels $s_0+1\dots s_0+s_1$,
the root child vertex with label  $i$ has $s_i$ children with labels $s_0+s_1+\dots s_{i-1} +1, \dots, s_0+s_1+\dots s_{i-1}+s_i$, and so on.
We call this labeling of $t$ the {\it natural labeling}
of a plane rooted tree $t$.
 So, using this labeling for $t$
we  have  that the set of vertices $V_t$ in a  plane rooted tree $t\in \TT^0_n$ admits a natural total order $\prec$, which we call the {\it plane-tree order}.
I.e., given two (distinct) vertices $u$ e $v$  of $t$, we have
$v\prec u$, and say that {\it $v$ is older than $u$} or {\it $u$ is younger than $v$}, if the natural label of $v$ is less than the natural label of $u$. In other words $v\prec u$ if either $d(v)<d(u)$,
or $d(v)=d(u)$ but $v$ is above $u$ in the drawing of $t$.
Given plane rooted tree $t\in \TT^0_n$ we will denote by $\th_t$ the unique labeled tree in $T^0_n$ whose labels coincides in all vertices with the natural labels of $t$, i.e. $\th_t\equiv \theta(t)$.

\vv
Let us further remark that the total order introduced on the vertices of a plane rooted tree $t\in \TT^0_n$ (via the natural labeling) automatically
induces a total order, still denoted by $\prec$,
also on vertices of a labeled rooted tree $\t\in T^0_n$.
Indeed given any two vertices $u,v$ in $\t\in T^0_n$ we say that $u\prec v$ if the corresponding vertices $m(u)$, $m(v)$ in $t=m(\t)\in \TT^0_n$
are such that $m(u)\prec m(v)$. Please note that this induced total order of the vertices of a labeled rooted tree, which, we recall,
are  integers numbers, can be different from  the standard order of the integers. Indeed if $u$ and $v$ are siblings in $t$ we have clearly that  $u<v$  implies $u\prec v$.
However, if $u$ and $v$ are not siblings then
it may well happens that $u<v$ but $v\prec u$.

\subsection{\it The Penrose map}

We recall that if   $\th \in T^0_n$ is a labeled rooted tree, then $V_\th$ is the vertex set of
$\th$ and  $E_\th$ is  the   edge set of $\th$. We also recall once again that  $V_\th\equiv\I^n_0$, i.e. the vertices of $\th$
are the integers $\{0,1,\dots,n\}$  with $0$ being the root. For a vertex $i\in V_\th$,
we recall that $d(i)$ denotes  the depth  of the vertex $i$ (i.e. its edge distance  from $0$)
and that $i^*$  denotes the parent of $i$.
Let further recall that a labeled tree $\th\in T^0_n$ can be viewed as a $I^0_n$-dressed tree $\th=(t,\s)$ where $t$ is a (unlabeled)
rooted tree with vertex set $V_t$ and root $r$
and $\s:V_t\to  \I_n^0$ is a bijection such that $\s(r)=0$. This  leads also to the observation that the pair
 $(\th; (x_0,x_1,\dots, x_n))$, where $\th=(t,\s)$ is a labeled rooted tree  (with $t$ rooted tree and $V_t\to \I^0_n$ bijection) and
 $(x_0,x_1, \dots, x_n)\in X^{n+1}$ is an ordered $n+1$-tuple, uniquely
determine a $X$-dressed tree $\t=(t,\tilde\s\circ\s)$
where $\tilde\s:\I^0_n\to X$ such that $\tilde\s(i)=x_i$.
Finally let us recall that the set of  vertices $V_\th$ in a labeled rooted tree $\th$    is equipped with the total order $\prec$
(induced by the underlying plane rooted tree $m(\th)$ ) previously defined.

Let's now go back to the graph $G=(X,E)$ in which the hard core lattice gas has been defined. We recall that if $x,y\in X$ are such that either $\{x,y\}\in E$ or $x=y$,
we denote this shortly with the symbol $x\not\sim y$ ($x$ and $y$ are incompatible  or $x$ and $y$ overlap), and
if $\{x,y\}\not\in E$, we denote shortly $x\sim y$ ($x$ and $y$  compatible).
We also recall that, for fixed $(x_0,x_1,\dots,x_n)\in X^{n+1}$,  $g(x_0,x_1,\dots,x_n)$ is the graph with vertex set $\I^0_n$ and edge set
 $E_{g(x_0,x_1,\dots,x_n)}=\{\{i,j\}\subset \I_0^n:\; x_i\nsim x_j\}$.

\begin{definition}\label{def0} The pair $(\th; (x_0,x_1,\dots, x_n))$ where $\th\in T^0_n$ and $(x_0,x_1,\dots, x_n)\in X^{n+1}$ is called a Penrose tree
if the following holds.

\begin{enumerate}

\item[{\rm (t0)}]   if $\{i,j\}\in E_\th$ then $\{i,j\}\in E_{g(x_0,x_1,\dots,x_n)}\Longleftrightarrow$ $  x_i\nsim x_j$


\item[{\rm (t1)}]  if two vertices $i$ and $j$ are such that $d(i)=d(j)$, then
$\{i,j\}\not\in  E_{g(x_0,x_1,\dots,x_n)}\Longleftrightarrow$ $  x_i\sim x_j$;

\item[{\rm (t2)}]  if two vertices $i$ and $j$ are such that $d(j)=d(i)-1$ and $i^*\prec j$, then
$\{i,j\}\not\in  E_{g(x_0,x_1,\dots,x_n)}$ (i.e.  $  x_i\sim x_j$).
\end{enumerate}
We denote by
$P{(x_0,x_1,\dots,x_n)}$ the subset  of $T^0_n$ constituted by those $\th\in T^0_n$ such that the pair $(\th; (x_0,x_1,\dots, x_n))$ is Penrose.
\end{definition}

\\{\bf Remark}. Property (t0) says that (labels of) children always overlap (labels of) their parents, property (t1)
says that siblings and/or cousins do not overlap. Finally property (t3) says that
children are always compatible with their uncles which  are younger  than the father (i.e. are below the father in the drawing of the plane tree). We want to emphasize
that the map presented  above  is slightly different respect to the original map given by Penrose in \cite{P} (used also
in \cite{FP}, \cite{So}, \cite{JPS}).  The present definition has the advantage to be  independent of the (integer) labels of the tree $\th\in T_n^0$.
It depends only on the underlying
plane rooted tree $t=m(\th)$ (since in condition t2 we are using the order $i^*\prec j$ which depends
only on the underlying plane rooted tree $m(\th)$ in place of the usual order $i^*< j$ used  in the original Penrose paper \cite{P} and in works \cite{FP,JPS,So}
which instead depends on the labels of $\th$). This  fact will be crucial in order to rewrite the series for \equ(TP)
for $\Pi_{x_0}(-\bm \r)$ in terms of plane rooted trees. 
\vskip.3cm

\begin{proposition}\label{pro9}

$$
\phi^{T}(x_0,x_{1},\dots ,x_{n})\; =\;(-1)^n
\sum\limits_{\th\in T_n^0} \1_{\th\in P(x_0,x_1,\dots,x_n)}(\th)
\Eq(r.20)
$$
where $\1_{\th\in P(x_0,x_1,\dots,x_n)}$ is the characteristic function of the set $P(x_0,x_1,\dots,x_n)$ in $T^0_n$, i.e.
$$
\1_{\th\in P(x_0,x_1,\dots,x_n)}(\th)\;=\;\cases{1 &if $\th\in P(x_0,x_1,\dots,x_n)$\cr\cr
0 & otherwise}
$$
\end{proposition}

\\{\bf Proof}. Fix $(x_0,x_1,\dots,x_n)\in X^{n+1}$.  Then is uniquely defined the
$g(x_0,x_1,\dots,x_n)$ with vertex set $\I^0_n$ and edge set
 $E_{g(x_0,x_1,\dots,x_n)}=\{\{i,j\}\subset \I_0^n:\; x_i\nsim x_j\}$. Without loss in generality we may assume that $g(x_0,x_1,\dots,x_n)$ is connected (otherwise
$\phi^{T}(x_0,x_{1},\dots ,x_{n})=0$ and
 \equ(r.20) is trivial). We denote by  ${G}^0_n$ the set of all connected graphs with vertex set $I^0_n$ and we put
 $$
G_{g(x_0,x_1,\dots,x_n)}= \{g\in {G}^0_n: g\subset  g(x_0,x_1,\dots,x_n)\}
$$
and
$$
T_{g(x_0,x_1,\dots,x_n)}=\{\th\in T^0_n: \th\subset g(x_0,x_1,\dots,x_n)\}
$$

Let us define the map $q: G_{g(x_0,x_1,\dots,x_n)}\to T_{g(x_0,x_1,\dots,x_n)}$ that associate to  $g\in  G_{g(x_0,x_1,\dots,x_n)}$
a unique labeled rooted tree $q(g)\in T_{ g(\g_0,\g_{1},\dots ,\g_{n})}$ as follows. We recall that the
vertices of $g\in G_{g(x_0,x_1,\dots,x_n)}$ are labeled with labels in $\{0,1,2,\dots, n\}$
and  we are denoting by $E_g$ the edge set of $g$. We also consider the graph $g$ as always rooted
in $0$, so for any $j$ vertex of $g$, we will denote by $d_g(j)$ its distance from the root $0$ in $g$.

1) We  first delete all edges $\{i,j\}$ in $E_g$
with $d_g(i)=d_g(j)$.  After this operation we are left with a connected graph $g'$ such that $d_{g'}(i)=d_{g}(i)$ for all vertices $i=0,1,\dots,n$.
Moreover
each edge $\{i,j\}$ of $g'$ is such that $|d_{g'}(i)-d_{g'}(j)|=1$.

2) Let $i_1,\dots, i_{s_0}$ be the vertices at distance 1 from the root $0$ in $g'$ ordered in such way that $i_1<i_2<\cdots< i_{s_0}$
(note that we identify vertices with their labels, so that $\{i_1,\dots, i_{s_0}\}$  is a subset
$\{0,1,2,\dots, n\}$). Now take the smaller of these vertices, say  $i_1$, and  let $j^{i_1}_1, \dots, j^{i_1}_{s_{i_1}}$
be the vertices connected to $i_1$ by edges of $E_{g'}$  (these vertices are at distance 2
 from the root $0$ and again are ordered according their labels) and delete all edges of $g'$ connecting vertices $j^{i_1}_1, \dots, j^{i_1}_{s_{i_1}}$
 to vertices in the set $\{i_2,\dots, i_{s_0}\}$. The graph so obtained
$g_1'$ is such that any of the vertices $j^{i_1}_1, \dots, j^{i_1}_{s_{i_1}}$ is connected
only to $i_1$ and vertices at distance greater than 2. Then take the vertex $i_2$ (the smaller after $i_1$)
and let  $j^{i_2}_1, \dots, j^{i_2}_{s_{i_2}}$ be the vertices connected to $i_2$ at distance 2 from the root $0$ in $g'_1$
and delete all edges of $g'_1$ connecting vertices $j^{i_2}_1, \dots, j^{i_2}_{s_{i_2}}$ to vertices in
the set $\{i_3,\dots, i_{s_0}\}$. The graph so obtained
$g_2'$ is such that any of the vertices $j^{i_1}_1, \dots, j^{i_1}_{s_{i_1}}$ is connected only to $i_2$ and
vertices at distance greater than 2. After $s_0$ steps we are left with a graph
$g'_{s_0}$ with no loops among vertices at distance $d\le 2$ from the root. Continue now this procedure until all
vertices of $g$ are exhausted, always respecting the order of the labels. Namely,
take $j^{i_1}_1$  (i.e. the one with the smaller label among $j^{i_1}_1, \dots, j^{i_1}_{s_{i_1}}$) and consider
the vertices at distance $3$ emanating from  $j^{i_1}_1$  and delete all edges
linking these vertices to some vertex in the set
$\{j^{i_1}_2, \dots, j^{i_1}_{s_{i_1}}, j^{i_2}_1, \dots, j^{i_2}_{s_{i_2}},\dots, j^{i_{s_0}}_1, \dots, j^{i_2}_{s_{i_{s_0}}}\}$ and continue this
procedure until all vertices are exhausted.
The resulting graph $g''\doteq q(g)$ is by construction  a spanning connected subgraph of $g(x_0,x_1,\ldots,x_n)$, i.e.
$q(g)\in  G_{g(x_0,x_1,\ldots,x_n)}$, and which has
no cycles, i.e. $q(g)\in T_{g(x_0,x_1,\dots,x_n)}$. Observe that the map $q$ is a surjection from $G_{g(x_0,x_1,\dots,x_n)}$ to $T_{g(x_0,x_1,\dots,x_n)}$.

Conversely, Let $p$ be the map that to each tree $\th \in T_{G(x_0,x_1,\dots,x_n)}$ associates the graph
$ p(\th)\in G_{g(x_0,x_1,\dots,x_n)}$ formed by adding  to $\th$
all edges $\{i,j\}\in E_{g(x_0,x_1,\dots,x_n)}\setminus E_\th$ such that
either $d_\th(i)=d_\th(j)$, or $d_\th(j)=d_\th(i)-1$ and $i^*\prec j$.

Observe now that the set  $G_{g(x_0,x_1,\dots,x_n)}$ is partially ordered by edge inclusion, namely, $g,g\in G_{g(x_0,x_1,\dots,x_n)}$
and $E_g\subset E_{g'}$, then $g<g'$. Moreover
if $g,g'\in G_{g(x_0,x_1,\dots,x_n)}$ and $g< g'$ we denote by $[g,g']$ the subset of $G_{g(x_0,x_1,\dots,x_n)}$ formed by those $\hat g$
such that $g<\hat g< g'$. With these definitions
we have that if $\th\in T_{g(x_0,x_1,\dots,x_n)}$ and $g\in [\th, p(\th)]$, then, by construction of the map $m$,
we have that  $m(g)=\th$,
i.e., among those graphs $g\in G_{g(x_0,x_{1},\dots ,x_{n})}$
such that $q(g)=\th$ $\th$ is the minimal graph and $p(\th)$ is the maximal graph, respect to the partial order relation $<$ in $G_{g(x_0,x_1,\dots,x_n)}$.
So $G_{g(x_0,x_1,\dots,x_n)}$ is partitioned in the disjoint union of the sets $[\th, p(\th)]$ with $\th\in T_{g(x_0,x_1,\dots,x_n)}$. This shows that
 the map $p$ provides a so-called {\it partition scheme} of the family
of graphs $G_{g(x_0,x_1,\dots,x_n)}$.
Observe finally, recalling Definition \ref{def0}, that if  $\th\in T_{g(x_0,x_1,\dots,x_n)}$, then
$p(\th)=\th \iff \th \in  P(x_0,x_1,\dots, x_n)$.

With these definition we have

$$
\sum\limits_{g\in  G_{g(x_0,x_1,\ldots,x_n)}}
 (-1)^{\card{E_g}}
 =
 \sum_{\th\in T_{ g(x_0,x_{1},\dots ,x_{n})}}(-1)^{|E_\th|}
 \sum_{g\in G_{g(x_0,x_1,\ldots,x_n)}\atop q(g)=\th}(-1)^{|E_g|-|E_\th|}=
$$
$$
= (-1)^{n}
\sum_{\th\in T_{ g(x_0,x_{1},\dots ,x_{n})}}
 [1+(-1)]^{|E_{p(\th)}|-|E_\th|}= (-1)^{n} \sum_{\th\in T_{ g(x_0,x_{1},\dots ,x_{n})}\atop p(\th)=\th }1=
 $$

$$
= (-1)^{n} \sum\limits_{\th\in T_n^0}
\1_{\th\in P(x_0,x_1,\dots,x_n)}
$$
and the proposition is proved. $\Box$
\vskip.2cm
\\Using this proposition we can rewrite the formal series \equ(TP) as

$$
\Pi_{x_0}(-\bm \r) \;=\;\sum_{n=0}^{\infty}{1\over n!} \sum\limits_{\th\in T_n^0}  \sum_{(x_1,\dots,x_n)\in X^n}
\1_{\th\in P(x_0,x_1,\dots,x_n)}\,\,{\r_{x_1}}\dots{\r_{x_n}}\; =
$$
$$
=\;\sum_{n=0}^{\infty}{1\over n!} \sum\limits_{\th\in T_n^0}  \phi_{x_0}(\th,\bm\r) \;\; \;\;
\,\,\,\,\,\,\,\,\,\,\,\,\,\,\,\,\,\,\,\,\,
\,\,\,\,\,\,\,\,\,\,\,\,\,\,\,\,\,\,\,\,\,\,\,\,\,\,\,\,\,\,\,\,\,\,\,\,\,\,\,
\,\,\,\,\Eq(pitre)
$$
where
$$
\phi_{x_0}(\th,\bm \r)=\; \sum_{(x_1,\dots,x_n)\in X^n}
\1_{\th\in P(x_0,x_1,\dots,x_n)}\,\,{\r_{x_1}}\dots{\r_{x_n}}\;\Eq(lbdp)
$$
This equation shows that the formal series $\Pi_{x_0}(-\bm \r)$ can be
reorganized as a sum over terms associated to labeled rooted trees.
Now, as remarked above, the factor $\phi_{x_0}(\th,\bm \r)$ defined in  \equ(lbdp) does not depend
on the labels of $\th\in T^0_n$ (the variables $x_1,\dots, x_n$ are mute variables) but only  on the plane rooted tree associated to $\th$ by the map
$m$ defined above. So
we can write, for any given $\th\in T^0_n$ such that $m(\th)=t\in \TT^0_n$
$$
\phi_{x_0}(\th,\bm\r)=  \phi_{x_0}(\th_t,\bm\r)\Eq(explw1)
$$
where, recall that $\th_t$ denotes the natural labeled tree associated to
$t$ (i.e. $t$ plus the natural labeling of the vertices according to the natural order defined before). Therefore
$$
\phi_{x_0}(\th,\bm\r)=\sum_{(x_1,\dots,x_n)\in X^n}
\1_{\th_t\in P(x_0,x_1,\dots,x_n)}\,\,{\r_{x_1}}\dots{\r_{x_n}}\;\Eq(lbdp2)
$$
and

$$
\Pi_{x_0}(-\bm \r)  =
\sum_{n=0}^{\infty}{1\over n!} \sum\limits_{\th\in T_n^0}  \phi_{x_0}(\th,\bm\r) \;=
\sum_{n\geq 0}{1\over n!}\sum\limits_{t\in \TT^0_{n}}\sum_{\th\in T^0_n\atop m(\th)=t}\phi_{x_0}(\th_t,\bm\r)=
$$
$$
=
\sum_{n\geq 0}{1\over n!}\sum\limits_{t\in \TT^0_{n}}\phi_{x_0}(\th_t,\bm\r)\sum_{\th\in T^0_n\atop m(\th)=t}1=
\sum_{n\geq 0}{1\over n!}\sum\limits_{t\in \TT^0_{n}}\phi_{x_0}(\th_t,\bm\r)|m^{-1}(t)|=
$$
$$
= \sum_{n\geq 0}\sum\limits_{t\in \TT^0_{n}}\Big[\prod_{v\in V_t}{1\over s_v!}\Big]\sum_{(x_1,\dots,x_n)\in X^n}
\1_{\th_t\in P(x_0,x_1,\dots,x_n)}\,\prod_{i=1}^n{\r_{x_i}}
$$
I.e. we have  obtained
$$
\Pi_{x_0}(-\bm \r) =  \sum_{n\geq 0}\sum\limits_{t\in \TT^0_{n}}\Big[\prod_{v\in V_t}{1\over s_v!}\Big]\sum_{(x_1,\dots,x_n)\in X^n}
\1_{\th_t\in P(x_0,x_1,\dots,x_n)}\,\prod_{i=1}^n{\r_{x_i}} \Eq(forma)
$$

\section{Witness trees are Penrose}

Let us now go back to the Moser Tardos scheme illustrated in section 1.4. We will make use of the concept on plane rooted tree previously introduced
to redefine the witness trees in a completely deterministic way.
From now on we suppose that the finite set $X$ which indexes the family of events $\bm A$  is ordered and indicate with $<$ such an order.
Let now $t\in \mathbb{T}^0$ be a plane rooted tree with vertex set $V_t$ and edge set $E_t$ and let $\s: V_t\to X$ be a function.
We say that $\s$
is a {\it good labeling of $t$}  if it is such
that $\{v,v'\}\in  E_t\iff \s(v)\not\sim \s(v')$ and moreover  if $v$ and $w$ are siblings and $v\prec w$ (in the natural order of the vertices of $t$) then $\s(v)<\s(w)$ (in the order introduced in $X$).

\begin{definition}\label{def2}
A  witness tree $\t=(t,\s)$ is a finite plane rooted tree  $t=(V_t,E_t)\in {\TT}_0$ together with a good labeling $\s: V_t\to X$.
\end{definition}\label{def1}

\\{\bf Remark}. Note that the definition of {\it proper} witness tree here above is  perfectly equivalent to the
Moser Tardos Definition \ref{pippo} given in section 1.4.
Indeed it is obvious that $\{v,v'\}\in  E_t\iff \s(v)\not\sim \s(v')$ is the same as requiring
that the children of a vertex  $u\in V_t$ receive labels from $\G^*(\s(u))$. Moreover, since labels of siblings must
respect their order in the plane tree, these labels must be necessarily distinct.
It is finally simple to construct  a one-to one correspondence
between rooted trees $ t$ whose vertices are labeled by a function $\s:V_t\to X$ according to the rule that
children always overlap their parents  and always receive distinct labels and
plane rooted trees $t$ whose vertices are good-labeled with labels from $X$. Indeed, any plane rooted tree $t$
whose vertices are good-labeled with labels from $X$ can also be viewed as  a  rooted trees $t$ whose vertices are labeled with labels from $X$.  Conversely, to any
rooted tree $t$ whose vertices are labeled with labels from $X$  according to the rule that the labels of the
children always overlap the labels of their parents  and always receive distinct labels we can associate a unique plane rooted tree $t$ whose vertices
are good-labeled with labels from $X$: just order the children
of the rooted tree $t$
according to the order of their labels in $X$ obtaining in this way a (unique) plane rooted tree $t$ whose vertices
are automatically good-labeled with labels from $X$.

\vskip.2cm

\begin{definition}\label{def3}
A proper witness tree  $\t=(t,\s)$  is called a  Penrose tree if the following occurs:
\begin{itemize}
 \item[{\rm (t1)}]  if two vertices $v$ and $v'$ are such that $d(v)=d(v')$, then $\s(v)\sim\s(v')$;

\item[{\rm (t3)}]  if two vertices $v$ and $v'$ are such that $d(v')=d(v)-1$ and $v^*\prec v'$  (i.e. $v'$ is an uncle of $v$ which is below  the father
$v^*$ of $v$), then $\s(v)\sim\s(v')$

\end{itemize}
We denote by ${\cal S}^x_X$ the set of all Penrose trees $\t=(t,\s)$ with root label $x$.
\end{definition}
\\{\bf Remark}. Note that this definition coincides, {\it mutatis mutandis} with definition \ref{def0} given in section 2 in the following sense. If $\t=(t,\s)$
is a Penrose tree according to definition \ref{def3}, then $t$, being a plane rooted tree, defines uniquely the labeled rooted
tree $\th_t\in T^0_n$ previously seen. Moreover the function $\s$ defines uniquely a $n+1$-tuple $(x_0,x_1,\dots, x_n)$
such that $\s(i)=x_i$ for each $i\in I^0_n$ (we are identifying vertices of $V_t$ with numbers in $I^0_n$
through the bijection $t\mapsto \th_t$). Then $\th_t\in P(x_0,x_1,\dots,x_n)$ according to definition \ref{def0}.
\vv

Now we are in the position to explain how Moser and Tardos  associate to each step $s$ of the algorithm, with log $C$,  a witness  tree $\t(s)\in T_X^{C(s)}$
(with vertex labels chosen in the set $X$ and root with label $C(s)$). The tree $\t(s)$ is obtained
by constructing a sequence
$\t_s(s), \t_{s-1}(s), \dots ,\t_{1}(s)$  of witness trees and then posing   $\t(s)= \t_1(s)$.
Let $\t_s(s)$
be the witness tree formed only by a single vertex (i.e. the root) with label $C(s)$. For $i-1\in \{1,\dots, s-1\}$,
$\t_{i-1}(s)$ is obtained from $\t_i(s)$ by attaching  a new vertex to $\t_i(s)$  with label $C(i-1)$ in the following way.
Let
 $W_i$ be constituted by all vertices of $\t_{i}(s)$ whose labels (which, recall,  are elements of $X$) are incompatible with the event $C(i-1)$.
If $W_i$ is empty  (i.e. all vertices in $\t_{i}(s)$ have labels in $X$ which are compatible with $C(i-1)$)
do nothing, i.e. put $\t_{i-1}(s)=\t_{i}(s)$  and skip to the next step.
If $W_i$ is not empty, then, quoting Moser and Tardos,
``choose among all such vertices the one  (say $v$) having the
maximum distance from the root and attach a new child vertex $u$ to $v$ with label $C(i-1)$, thereby obtaining
the tree $\t_{i-1}(s)$". Of course, even if not stated explicitly, if there is more than one vertex in $W_i$ at maximum distance from the root, then one has to choose
one, at random or according to some deterministic  rule.
Moser and Tardos  does not give any deterministic rule to choose the vertex $v$ when a choice is necessary. That is, if $v$ is non unique, they choose it at random. We instead give
a deterministic rule (that's why we need to work with plane rooted trees in place of simple rooted trees).
Let $\tilde W_i$ be the subset of $W_i$ formed by those vertices of $W_i$ which are at the maximal
distance from the root \footnote{Be careful! Here 'maximal' means maximal in $ W_i$, so that $\t^{(i)}(s)$ can have vertices with depth greater than those in $\tilde W_i$
as long as  the labels in these vertices are all compatible with $C(i-1)$} and attach the  new vertex $u$ with
label $C(i-1)$ to the lowest (a.k.a. younger) vertex of the set $\tilde W_i$, say $v$,  according to the tree order
described before, forming in this way the tree $\t_{i-1}(s)$, in which $u$ is a child of $v$. Of course, in order
to obtain  a good labeling of $\t_{i-1}(s)$,  if the vertex $v$ had already children in  $\t_{i}(s)$
(so that $u$ becomes a new sibling of these children of $v$)
attach the new vertex $u$ with label $C(i-1)$ respecting the order of the  children of $v$.

\vskip.2cm

According to Moser and Tardos we say that a witness tree $\t$ occurs in the log $C$ of the algorithm
if there exists $s\in \mathbb{N}$ such that $\t(s)=\t$. We now prove
a generalization of Lemma 2.1 in \cite{MT}. Recall that $\T_X^x$ denotes the set of all distinct witness tree
which can be  generated by the algorithm according to the procedure described above and with root labeled $x\in X$.

\begin{proposition}\label{propo} Let $\t=(t,\s)$ be  a proper witness tree and let $C$ be the (random) log
produced by the algorithm. If $\t$ occurs in the log $C$,  then $\t$ is a Penrose tree.
\end{proposition}
\\{\it Proof}. If $\t$ occurs in the log $C$, then, by definition,  there exists $s\in \mathbb{N}$ such that $\t(s)=\t=(t,\s)$.
By construction the plane rooted tree $t$
associated to  $\t$ is such that in any given vertex $v\in V_t$ the label $\s(v)$ is compatible with all labels at the same distance from the root. Indeed, suppose  by absurd,
that $v$ and  $v'$ are two vertices of $\t(s)=\t$ at the same distance from the root, i.e. $d(v)=d(v')$ and that the label of $v$ is incompatible with the label of $v'$.
Suppose, without loss in generality, that $v'$ has been attached after $v$. But then, since  the label
of $v'$ is incompatible with the label of $v$, we have that  $d(v')\ge d(v)+1$ contrary to the hypothesis that  $d(v')=d(v)$.
So if $d(v)=d(v')$
then necessarily $\s(v)\sim \s(v')$.

Suppose now that $v$ and $v'$ are vertices of $\t$ such that $v'$ is an uncle of $v$ who is younger than the father $v^*$ of $v$
(i.e. an uncle of $v$ which is below the father $v^*$ of $v$ in the drawing of $\t$). Then we have to show that
$\s(v)\sim \s(v')$. Indeed, suppose by absurd that $v'$ is a younger uncle of $v$ and that $\s(v)\not\sim \s(v')$. We have to consider two cases.
First we suppose that $v$ has been added after $v'$ to form $\t(s)=\t$.
Since $\s(v)\not\sim \s(v')$ and $v'$ is below $v^*$, then, according to the deterministic rule described above,
$v$ cannot be attached to $v^*$: it must be attached to $v'$ or to another uncle below $v'$,
contrary to the hypothesis that $v$ is attached to $v^*$. Secondly, suppose that $v'$ has
been added after $v$. But then $d(v')\ge d(v)+1$, contrary to the hypothesis that $v'$ is uncle of $v$ (and hence $d(v')=d(v)-1$). $\Box$
\vv
Recalling the definition of $T^x_X$  given in section 1.4 and the definition of  the set ${\cal S}^x_X$, we can resume
Proposition \ref{propo} by saying that $T^x_X\subset {\cal S}^x_X$. Therefore recalling formulas \equ(expn) and \equ(77)
we have that the expected number of times an event $A_x\in \bm A$ is resampled by the
MT-algorithm is bounded by
$$
E(N_x)\le \tilde\Phi_x(\bm p)\Eq(prus)
$$
where
$$
 \tilde\Phi_x(\bm p)=\sum_{\t=(t,\s)\in {\cal S}^x_X}\prod_{v\in V_t}p_{\s(v)}\Eq(778)
$$
Now note that the sum on the r.h.s. of \equ(778) can also be rewritten as follows. Let us denote by ${\cal S}^{x,n}_X$
the set of Penrose trees  with $n+1$ vertices (the root plus $n$ vertices) with root $r$ carrying the label $x\in X$. Then
$$
 \tilde\Phi_x(\bm p)= p_x\sum_{n=0}^\infty\sum_{\t \in {\cal S}^{x,n}_X} \prod_{v\in V_t\atop v\neq r}^n p_{\s(v)}
$$
Now recall that a Penrose  tree is a witness tree i.e.  a pair $\t=(t,\s)$ where $t$ is a plane rooted tree and $\s$ is a good labeling of the vertices of $t$. Moreover
we recall also that the vertices of a plane rooted tree are naturally labeled by integers according to the order $\prec$ seen above. Hence a good labeling $\s$
can be also denoted by an $n$-tuple $\s=(x_1,\dots, x_n)$ such that $x_i$ is the label attached to $i^{\rm th}$ vertex of $t$ (in the order $\prec$). Therefore
$$
\sum_{\t \in {\cal S}^{x,n}_X} \prod_{v\in V_t\atop v\neq r}^n p_{\s(v)})=\sum_{t\in \mathbb{T}_n^0}\sum_{(x_1,\dots, x_n)\in X^n\atop {\rm good~labeling}}  \prod_{i=1}^n p_{x_i}
$$
Now, to say that $(x_1,\dots, x_n)$ is a good labeling for a given $t\in \mathbb{T}_n^0$ means two things: 1) the set $X$
is ordered and the labels in $X$ attached to the vertices of $t$
must respect the order the children at each vertex  (ordered labeling); 2) the labels in $X$ must respect the Penrose rule of the plane rooted tree,
i.e. $\s=(x_1,\dots, x_n)$ must be such that
$\t=(t,\s)$ is Penrose according to the definition \ref{def3}. As observed above, this is the same as requiring that, for fixed $\s=(x,x_1,\dots,x_n)\in X^{n+1}$,
the  labeled  tree $\th_t$ uniquely associated to $t$ by the map
$\theta$ seen before is Penrose according
to the definition \ref{def0}. I.e. $\th_t\in P(x_0,x_1,\dots,x_n)$. Therefore, recalling \equ(forma), we get
$$
\sum_{t\in \mathbb{T}_n^0}\sum_{(x_1,\dots, x_n)\in  X^n\atop {\rm good~labeling}}  \prod_{i=1}^n p_{x_i}=
\sum_{t\in \mathbb{T}_n^0}\sum_{(x_1,\dots, x_n)\in  X^n\atop {\rm ordered~labeling}}
\1_{\th_t\in P(x_0,x_1,\dots,x_n)} \prod_{i=1}^n p_{x_i}=
$$
$$
= \sum_{t\in \mathbb{T}_n^0}\Big[\prod_{v\in V_t}{1\over s_v!}\Big]\sum_{(x_1,\dots, x_n)\in X^n}
\1_{\th_t\in P(x_0,x_1,\dots,x_n)} \prod_{i=1}^n p_{x_i}= \Pi_{x_0}(-\bm p)
$$
Hence, recalling definition \equ (778) and bound \equ(prus), we conclude that
$$
E(N_x) \le \tilde \Phi_x(\bm p)= p_x\Pi_x(-\bm p)
$$
which concludes the proof of Theorem \ref{MTn}.

\section*{Acknowledgments}
Aldo Procacci has   been partially supported by the Brazilian  agencies
Conselho Nacional de Desenvolvimento Cient\'{\i}fico e Tecnol\'ogico
(CNPq)  and  Funda{\c{c}}\~ao de Amparo \`a  Pesquisa do estado de Minas Gerais (FAPEMIG - Programa de Pesquisador Mineiro).

\end{document}